\documentclass[seceq]{ptptex}
\usepackage[dvips]{graphicx}

\pubinfo{Vol.~109, No.~2, February 2003}
\notypesetlogo                       %comment in if to eliminate PTPTeX 

\title{Does Sub-Barrier Bremsstrahlung in $\alpha$-Decay of
$^{210}$Po Exist?}
 
\author{%
Sergei P. \textsc{Maydanyuk}$^{}$\footnote{E-mail:
maidan@kinr.kiev.ua}
and Vladislav S. \textsc{Olkhovsky}$^{}$\footnote{E-mail:
olkhovsk@kinr.kiev.ua}
}
 
\inst{%
Institute of Nuclear Research NASU,
prosp. Nauki, 47, Kiev-28, 03680, Ukraine}

\recdate{%      
September 6, 2002}
\abst{
A % non-stationary
quantum mechanical model for the description of the
$\alpha$-decay of heavy nuclei with accompanying photons emission
is presented. The model is based on a quantum mechanical
one-particle model of $\alpha$-decay through a decay barrier.
The bremsstrahlung spectrum calculation employs  multipole
expansion of the vector potential of the Coulomb field of the
daughter nucleus and takes into account the dependence on the angle
between the directions of the $\alpha$-particle propagation and the
photon emission.
Spectra of $^{210}$Po are obtained for the angles
$25^{\circ}$ and $90^{\circ}$, and the best agreement with the
experimental data in the $90^{\circ}$ case %\cite{Eremin.2000.PRLTA}
in a comparison with other existing models is achieved.
From the angular analysis, the model gives monotonic
behavior of the bremsstrahlung spectrum for any value of the angle.
We find that sub-barrier photon emission (i.~e. bremsstrahlung photon
emission during $\alpha$-particle tunneling through decay barrier)
exists but gives a small contribution to the total bremsstrahlung
spectrum.
}

\begin{document}

\maketitle
% ***************************************************************************

% ***************************************************************************
%                               Introduction
% ***************************************************************************
\section{Introduction
\label{sec.1}}

The existence of sub-barrier bremsstrahlung phenomena in
$\alpha$-decay (i.e., brems\linebreak -strahlung photon emission during
tunneling of an $\alpha$-particle through a decay barrier) has not
yet been confirmed through the analysis of experimental data.
In Ref.~\citen{Maydanyuk.2002.IRAN.p1531} we proposed a quantum
mechanical approach to resolve this problem. With this approach, one
can calculate the bremsstrahlung spectra both with and without
sub-barrier photon emission for the nuclei $^{210}$Po, $^{214}$Po,
$^{226}$Ra and $^{244}$Cm, for which experimental data
\cite{Kasagi.1997.JPHGB,Kasagi.1997.PRLTA,Kasagi.2000.PRLTA,D'Arrigo.1994.PHLTA}
exist, and with a comparative analysis determine whether sub-barrier
bremsstrahlung exists.

This approach can be realized only with such a model that yields
sufficiently good agreement with experimental data. Note, that
the present theoretical description of experimental data
\cite{Kasagi.1997.JPHGB} for $^{210}$Po is unsatisfactory, and there
is a difficulty in  explaining the existence of the ``hole'' in the
bremsstrahlung spectrum. Note that these experimental spectrum
carried out for a value of the angle between the direction of the
photon emission and the direction of the $\alpha$-particle
propagation equal to $25^{\circ}$. New experimental data
\cite{Eremin.2000.PRLTA} of the bremsstrahlung spectrum for $^{210}$Po
have been obtained for the different value of the angle equal to
$90^{\circ}$, exhibit monotonic behaviour and have no ``hole'' with
such amplitude.
We suppose that obtaining a description of the bremsstrahlung spectrum
for $^{210}$Po with a sufficient degree of accuracy and clarification
of the spectrum behaviour are important problems  in the investigation
of sub-barrier bremsstrahlung phenomena.

To make possible a comparative analysis of these two groups of 
experimental data for $^{210}$Po obtained for different angular 
values, we present here an angular quantum mechanical model of the
$\alpha$-decay of heavy nuclei with accompanying (spontaneous)
emission of photons. The foundation of this work was laid in
Ref.~\citen{Maydanyuk.2001.UPJ},
where we calculated the bremsstrahlung spectrum for the nucleus
$^{214}$Po, taking into account E1.

This model is based on a quantum mechanical one-particle
model of $\alpha$-decay through a decay barrier \cite{Baz.1971}.
A description of the electro-magnetic field of the daughter
nucleus and the emission of photons in the $\alpha$-decay is
constructed on the basis of quantum electrodynamics \cite{Berestetski.1989}.
In calculating the bremsstrahlung spectrum, we use an expansion of
the vector potential of the electro-magnetic field of the daughter
nucleus in electric and magnetic multipoles,\cite{Aizenberg.1973}
in contrast to Papenbrock and Bertsch quantum mechanical
method,\cite{Papenbrock.1998.PRLTA} and we do not use the additional
transformation (\ref{eq.3.2.2}) used in Tkalya model\cite{Tkalya.1999.PHRVA}.
We first present  the mathematical tools of this model.
% ***************************************************************************

% ***************************************************************************
\section{A non-stationary quantum mechanical model of
$\alpha$-decay \\with bremsstrahlung}
\label{sec.3}

For investigation of the photon emission process, we consider the
state of the system (the $\alpha$-particle and the daughter nucleus)
before the emission as the initial ($i$-state) and the state of the
system after the emission as the final ($f$-state).
We calculate the bremsstrahlung spectrum using the transition
matrix element\cite{Landau.1989}
\begin{equation}
\begin{array}{ll}
  a_{fi} = \bigl< k_{f}, n_{k}+1 \bigl|
                \tilde{V}(\boldsymbol{r}, t)| k_{i}, n_{k} \bigr>, &
  \tilde{V} (\boldsymbol{r}, t) =
                -i \displaystyle\int\limits_{0}^{t}
                 e^{i\hat{H}_{0}t'}
                \hat{V}(\boldsymbol{r}, t') e^{-i\hat{H}_{0}t'} dt',
\end{array}
\label{eq.3.1.1}
\end{equation}                                          %       (3.1.1)
where
$\psi_{i}(\boldsymbol{r}) = |k_{i}\bigr>$ and
$\psi_{f}(\boldsymbol{r}) = |k_{f}\bigr>$ are
the % stationary
wave functions of the system in the initial
$i$-state and the final $f$-state from the unperturbed operator
$\hat{H}_{0}$ which describes scattering of the $\alpha$-particle
upon the daughter nucleus, uses the barrier and does not take
into account the photon emission;
$w = E$;
$n_{k}$ is the  number of photons of one sort with impulse
$\boldsymbol{k}$ in the initial $i$-state.
We choose units for which $\hbar = 1$, and $c = 1$.
In the Coulomb calibration, the interaction operator has the form
\cite{Aizenberg.1973,Landau.1989}
\begin{equation}
  \hat{V}(\boldsymbol{r}, t) = -Z_{\rm eff} \displaystyle\frac{e}{m}\boldsymbol{Ap},
\label{eq.3.1.2}
\end{equation}                                          %       (3.1.2)
where $Z_{\rm eff}$ is the effective charge,
$m$ is the reduced mass of the system, and 
$\boldsymbol{A}$ is the vector potential of the electromagnetic
field of the daughter nucleus.

We choose the vector potential $\boldsymbol{A}(\boldsymbol{r}, t)$
to take the form \cite{Berestetski.1989}
\begin{equation}
\begin{array}{ll}
  \boldsymbol{A}(\boldsymbol{r}, t) =
        \sum\limits_{\boldsymbol{k}, \alpha}
        \biggl(\hat{c}_{\boldsymbol{k}, \alpha} \boldsymbol{A}_{\boldsymbol{k}, \alpha} +
        \hat{c}^{+}_{\boldsymbol{k}, \alpha} \boldsymbol{A}^{*}_{\boldsymbol{k}, \alpha}
        \biggr), &
  \boldsymbol{A}_{\boldsymbol{k}, \alpha} =
        \sqrt{\displaystyle\frac{2\pi}{w}}
        \boldsymbol{e}^{(\alpha)}e^{i(\boldsymbol{kr}-wt)},
\end{array}
\label{eq.3.1.3}
\end{equation}                                          %       (3.1.3)
where $\boldsymbol{e}^{(\alpha)}$ is the unit polarization vector of
the photon, $\boldsymbol{k}$ is the wave vector of the photon, and
$w = k = \bigl| \boldsymbol{k}\bigr|$. The vector
$\boldsymbol{e}^{(\alpha)}$ is perpendicular to $\boldsymbol{k}$ in
the Coulomb calibration. We have two independent polarizations,
$\boldsymbol{e}^{(1)}$ and $\boldsymbol{e}^{(2)}$ for any photon
with impulse $\boldsymbol{k}$.

Let us consider a wave packet of the form
\begin{equation}
  \Psi_{i,f}(\boldsymbol{r}, t) =
  \int\limits_{0}^{+\infty} g(k - k_{i,f})
  \psi_{i,f}(\boldsymbol{r}, k) e^{-iw(k)t} dk,
\label{eq.3.1.4}
\end{equation}                                          %       (3.1.4)
where in the expressions $\Psi_{i,f}(\boldsymbol{r}, t)$ and
$k_{i,f}$ we use the index $i$ or $f$ only in dependence on 
consideration of the initial $i$- or final $f$-state of the system.
We take the weight amplitude $g(k - k_{i,f})$ in the form of a
Gaussian \cite{Maydanyuk.2000.UPJ}:
\begin{equation}
  g(k - k_{i,f}) = C \exp{\Biggl(-\displaystyle\frac
  {(k-k_{i,f})^{2}}{2(\Delta k)^{2}}\Biggr)}
\label{eq.3.1.5}
\end{equation}                                          %       (3.1.5)
where $C$ is a normalization factor and can be obtained from the
normalization condition $\int |g(k - k_{i,f})|^{2} dk = 1$.

The wave packets (\ref{eq.3.1.4}) can be used for analyzing in time
the tunneling of the $\alpha$-particle through the decay barrier and
allow to calculate the tunneling time of this particle in dependence
on the energy level of such process.
Preliminary calculations of these times of the $\alpha$-decay
with taking into account a possibility to emit photons from the
barrier region are presented in Ref.~\citen{Maydanyuk.2001.UPJ}.
Development of the model of the bremsstrahlung in the $\alpha$-decay
on the basis of the wave functions in form (\ref{eq.3.1.4}) allows
to calculate both the bremsstrahlung spectra and tunneling times.
Therefore, we use the wave packets (\ref{eq.3.1.4}) for
$\psi_{i}(\boldsymbol{r})$ and $\psi_{f}(\boldsymbol{r})$, with the
time factor $\exp{(-iwt)}$ \cite{D'Arrigo.1993.YAFIA}.

Substituting this wave packet into (\ref{eq.3.1.1}) and taking
account of the property
\begin{equation}
\begin{array}{ll}
  e^{-i\hat{H}_{0}t} \psi_{i}(\boldsymbol{r}) =
  e^{-iw_{i}t} \psi_{i}(\boldsymbol{r}), &
  \psi_{f}^{*}(\boldsymbol{r}) e^{i\hat{H}_{0}t} =
  \psi_{f}^{*}(\boldsymbol{r}) e^{iw_{f}t},         
\end{array}
\label{eq.3.1.6}
\end{equation}                                          %       (3.1.6)
we obtain the matrix element in the form
\begin{equation}
  \tilde{a}_{fi} =
  \int\limits_{0}^{+\infty} dk_{1}
  \int\limits_{0}^{+\infty} dk_{2}
  g(k_{1} - k_{i})g(k_{2} - k_{f})^{*}
  \bigl<k_{2}, n_{k}+1 \bigl| \tilde{V}(t) \bigr| k_{1}, n_{k} \bigr>,
\label{eq.3.1.7}
\end{equation}                                          %       (3.1.7)
where
\begin{equation}
  \tilde{V} (t) =
  -i \int\limits_{0}^{t} e^{iw_{2}t^{\prime}}
                \hat{V} e^{-iw_{1}t^{\prime}} dt^{\prime}.
\label{eq.3.1.8}
\end{equation}                                          %       (3.1.8)

We study the spontaneous emission of one photon with an impulse
$\boldsymbol{k}$. Therefore, $n_{k} = 0$, and the sum
$\sum\limits_{\boldsymbol{k}}$ can be omitted in the calculation
of the matrix element (\ref{eq.3.1.1}).

For calculation of the bremsstrahlung spectra we use the stationary
approximation: $t \to +\infty$.
Taking into account (\ref{eq.3.1.2}), (\ref{eq.3.1.3}), (\ref{eq.3.1.7})
and (\ref{eq.3.1.8}), and also using the definition of the
$\delta$-function, we calculate the transition matrix element
\begin{equation}
  \bigl< k_{2}, 1 \bigl| \tilde{V} \bigr| k_{1}, 0 \bigr> =
  F_{21} 2\pi\delta(w_{2} + w - w_{1}),
\label{eq.3.1.10}
\end{equation}                                          %       (3.1.10)
where
\begin{eqnarray}
  F_{21} & = & Z_{\rm eff}\displaystyle\frac{e}{m}
        \sqrt{\displaystyle\frac{2\pi}{w}} p(k_{1}, k_{2}),  \nonumber   \\
  p(k_{1}, k_{2})& = &
        \sum\limits_{\alpha = 1, 2}
        \boldsymbol{e}^{(\alpha) *}
        \biggl< k_{2} \biggl| e^{-i\boldsymbol{kr}}
        \displaystyle\frac{\bf\partial}{\partial  \boldsymbol{r}}
        \biggr| k_{1} \biggr>.
\label{eq.3.1.11}
\end{eqnarray}                                          %       (3.1.11)

For monochromatic particles, we obtain\cite{Berestetski.1989,D'Arrigo.1993.YAFIA}
\begin{equation}
  a_{fi} = 2\pi F_{fi} \bigl|C\bigr|^{2} \delta(w + w_{f} - w_{i}).
\label{eq.3.1.12}
\end{equation}                                          %       (3.1.12)
%
% where $C$ is a normalization factor ($\bigl|C\bigr|^{2} = 1$).

We define the transition probability from the initial $i$-state to
the final $f$-state in a unit time with a single photon emission of
impulse $\boldsymbol{k}$ and polarization $\boldsymbol{e}^{(\alpha)}$
as follows\cite{Berestetski.1989}:
\begin{eqnarray}
  \displaystyle\frac{dW}{d\Omega_{\nu}} & = &
        \displaystyle\frac{mk_{f}}{(2\pi)^{3}}
        \displaystyle\frac{w^{2}_{fi} \bigl| F_{fi} \bigr|^{2}}
        {(2\pi)^{2}} =
        \displaystyle\frac{Z^{2}_{\rm eff}e^{2}k_{f}w_{fi}}{(2\pi)^{4}m}
        \bigl| p(k_{i}, k_{f}) \bigr|^{2}, \nonumber \\
  w_{fi} & = & E_{i} - E_{f}.
\label{eq.3.1.13}
\end{eqnarray}                                          %       (3.1.13)
% ***************************************************************************

% ***************************************************************************
%       A multipole approach for the calculation of bremsstrahlung spectrum
% ***************************************************************************

The most difficult step in calculating the bremsstrahlung spectrum
is calculating $p(k_{i}, k_{f})$. We have
\begin{equation}
  p(k_{i}, k_{f}) =
        \sum\limits_{\alpha = 1, 2}
        \boldsymbol{e}^{(\alpha)*}
        \int\limits^{+\infty}_{0} dr \int d\Omega
        r^{2} \psi^{*}_{f}(\boldsymbol{r})
        e^{-i\boldsymbol{kr}}
        \displaystyle\frac{\partial}{\partial \boldsymbol{r}}
        \psi_{i}(\boldsymbol{r}).
\label{eq.3.2.1}
\end{equation}                                          %       (3.2.1)

In calculating $p(k_{i}, k_{f})$, various approaches can be used,
and the various quantum mechanical methods for these approaches differ.
Papenbrock and Bertsch used a dipole approximation in obtaining a
description of the vector potential of the electro-magnetic
field\cite{Papenbrock.1998.PRLTA}.
Tkalya used a multipole approach in calculating the bremsstrahlung
spectrum\cite{Tkalya.1999.PHRVA}, taking into account E1 and E2.
However, in calculating the integral (\ref{eq.3.2.1}) both in Tkalya
model and in Papenbrock and Bertsch model the transformation
\begin{equation}
  -\langle f | \boldsymbol{p} | i \rangle =
  \langle f |[H, \boldsymbol{p}]| i \rangle \displaystyle\frac{1}{w_{fi}} =
  i\hbar \biggl\langle f \biggl|
  \displaystyle\frac{\partial}{\partial \boldsymbol{r}} U(r) \biggr|
  i \biggl \rangle
  \displaystyle\frac{1}{w_{fi}}
\label{eq.3.2.2}
\end{equation}                                          %       (3.2.2)
is used. This transformation can be used only in a case that the 
approximation
\begin{equation}
  \exp(i\boldsymbol{kr}) = 1
\label{eq.3.2.3}
\end{equation}                                          %       (3.2.3)
is applied (see Appendix \ref{app.1}).
Therefore, using the transformation (\ref{eq.3.2.2}), the accuracy
of the calculation of the integral decreases, and the calculations
of the spectra with the models\cite{Papenbrock.1998.PRLTA,Tkalya.1999.PHRVA} are
carried out not in the dipole approximation or the multipolar
approximation (taking into account E1 and E2) but in the
approximation (\ref {eq.3.2.3}). For this  reason, we do not use
(\ref {eq.3.2.2}).
 
In computing the integral (\ref{eq.3.2.1}), we apply the multipolar
expansion of the vector potential of the electromagnetic field
without using the transformation (\ref{eq.3.2.2}). We write the
polarization vectors $\boldsymbol{e}^{\alpha}$ in terms of circular
polarization vectors ${\bf\xi}$ with opposite directions of
rotation\cite{Berestetski.1989}:
\begin{equation}
\begin{array}{ll}
  {\bf \xi}_{-1} = \displaystyle\frac{1}{\sqrt{2}}
                   (\boldsymbol{e}^{1} - i\boldsymbol{e}^{2}), &
  {\bf \xi}_{+1} = -\displaystyle\frac{1}{\sqrt{2}}
                   (\boldsymbol{e}^{1} + i\boldsymbol{e}^{2}).
\end{array}
\label{eq.3.2.4}
\end{equation}                                          %       (3.2.4)
We obtain
\begin{equation}
  p(k_{i}, k_{f}) =
        \sum\limits_{\mu = -1, 1}
        h_{\mu}{\bf \xi}^{*}_{\mu}
        \int\limits^{+\infty}_{0} dr \int d\Omega
        r^{2} \psi^{*}_{f}(\boldsymbol{r})
        e^{-i\boldsymbol{kr}}
        \displaystyle\frac{\partial}{\partial \boldsymbol{r}}
        \psi_{i}(\boldsymbol{r}),
\label{eq.3.2.5}
\end{equation}                                          %       (3.2.5)
where
\begin{equation}
\begin{array}{ll}
  h_{-1} = \displaystyle\frac{1}{\sqrt{2}} (1 - i), &
  h_{1}  = \displaystyle\frac{1}{\sqrt{2}} (-1 - i).
\end{array}
\label{eq.3.2.6}
\end{equation}                                          %       (3.2.6)

For the initial $i$- and final $f$-states, we can write
\begin{equation}
  \psi_{i,f}(\boldsymbol{r}) =
        \psi_{i,f}(r) Y_{lm}(\boldsymbol{n}^{i,f}_{r}),
\label{eq.3.2.7}
\end{equation}                                          %       (3.2.7)
where $Y_{lm}(\boldsymbol{n}^{i,f}_{r})$ are the normalized spherical
functions. Following Ref.~\citen{Aizenberg.1973}, we have
\begin{eqnarray}
%\begin{array}{lcl}
  \displaystyle\frac{\partial}{\partial \boldsymbol{r}}
  \psi_{i}(\boldsymbol{r}) & = &
        -\displaystyle\frac{d\psi_{i}(r)}{dr}
        \boldsymbol{T}_{01,0}(\boldsymbol{n}^{i}_{r}),       \nonumber  \\
  \boldsymbol{T}_{01,0}(\boldsymbol{n}^{i}_{r}) & = &
        \sum\limits^{1}_{\mu = -1} (110|-\mu\mu 0)
        Y_{1,-\mu}(\boldsymbol{n}^{i}_{r}) {\bf \xi}_{\mu},
%\end{array}
\label{eq.3.2.8}
\end{eqnarray}                                          %       (3.2.8)
where
$\boldsymbol{T}_{jl,m}$ are the vector spherical harmonics,
and $(j_{a}1j | -\mu\mu 0)$ are the Clebsch-Gordan coefficients.

We expand the vector potential $\boldsymbol{A} (\boldsymbol{r}, t)$
in multipoles as\cite{Aizenberg.1973}
\begin{equation}
  {\bf \xi}_{\mu}e^{i\boldsymbol{kr}} =
        \sqrt{2\pi}\mu \sum\limits_{l=1} \biggl(\sqrt{2l+1} i^{l}
        [\boldsymbol{A}_{lm}(\boldsymbol{r}, M) + i\mu\boldsymbol{A}_{lm}(\boldsymbol{r}, E)]
        \biggr),
\label{eq.3.2.9}
\end{equation}                                          %       (3.2.9)
with
\begin{equation}
\begin{array}{lcl}
  \boldsymbol{A}_{lm}(\boldsymbol{r}, M) & = &
        j_{l}(kr) \boldsymbol{T}_{ll,\mu}(\boldsymbol{n_{r}}),    \\
  \boldsymbol{A}_{lm}(\boldsymbol{r}, E) & = &
        \sqrt{\displaystyle\frac{l+1}{2l+1}}
        j_{l-1}(kr) \boldsymbol{T}_{ll-1,\mu}(\boldsymbol{n_{r}})-
        \sqrt{\displaystyle\frac{l}{2l+1}}
        j_{l+1}(kr) \boldsymbol{T}_{ll+1,\mu}(\boldsymbol{n_{r}}),
\end{array}
\label{eq.3.2.10}
\end{equation}                                          %       (3.2.10)
where
$j_{l}(kr)$ is the spherical Bessel function of order $l$,
and $\textbf{A}_{lm}(\boldsymbol{r}, M)$ and
$\textbf{A}_{lm}(\boldsymbol{r}, E)$ are the magnetic and electrical
multipoles, respectively.

From the calculations, we obtain $p(k_{i}, k_{f})$ as
\begin{eqnarray}
%\begin{array}{lcl}
  p(k_{i}, k_{f}) & = &
        \sqrt{2\pi} \sum\limits_{l=1} \biggl(\sqrt{2l+1} (-i)^{l}
        [p^{Ml}(k_{i}, k_{f}) - ip^{El}(k_{i}, k_{f})] \biggr), \nonumber \\
  p^{Ml}(k_{i}, k_{f}) & = &
        I_{1}J_{l}(l),                \nonumber                          \\
  p^{El}(k_{i}, k_{f}) & = &
        -\sqrt{\displaystyle\frac{l+1}{2l+1}} I_{2}J_{l}(l-1) +
         \sqrt{\displaystyle\frac{l}{2l+1}}   I_{3}J_{l}(l+1),
%\end{array}
\label{eq.3.2.11}
\end{eqnarray}                                          %       (3.2.11)
where
\begin{equation}
  J_{l}(n) = \int\limits^{+\infty}_{0}
        r^{2}\psi^{*}_{f,l}(r)
        \displaystyle\frac{d\psi_{i}(r)}{dr}
        j_{n}(kr) dr,
\label{eq.3.2.12}
\end{equation}                                          %       (3.2.12)
\begin{eqnarray}
  I_{1} & = &
        \sum\limits_{\mu=-1}^{1} \mu\hbar_{\mu} \int
        Y_{LM}^{*}(\boldsymbol{n}^{f}_{r})
        \boldsymbol{T}_{01,0}(\boldsymbol{n}^{i}_{r})
        \boldsymbol{T}_{ll,\mu}^{*}(\boldsymbol{n} _{\nu}) d\Omega,\nonumber 
       \\
  I_{2} & = &
        \sum\limits_{\mu=-1}^{1} \mu^{2}\hbar_{\mu} \int
        Y_{LM}^{*}(\boldsymbol{n}^{f}_{r})
        \boldsymbol{T}_{01,0}(\boldsymbol{n}^{i}_{r})
        \boldsymbol{T}_{ll-1,\mu}^{*}(\boldsymbol{n} _{\nu}) d\Omega,
        \nonumber \\
  I_{3} & = &
        \sum\limits_{\mu=-1}^{1} \mu^{2}\hbar_{\mu} \int
        Y_{LM}^{*}(\boldsymbol{n}^{f}_{r})
        \boldsymbol{T}_{01,0}(\boldsymbol{n}^{i}_{r})
        \boldsymbol{T}_{ll+1,\mu}^{*}(\boldsymbol{n}_{\nu}) d\Omega,
\label{eq.3.2.13}
\end{eqnarray}                                          %       (3.2.13)
and $n \in N$ (here $N$ is the natural number set),
$l$ is the quantum number for the wave function of the final
$f$-state. For the transitions M1 and E1 we obtain
\begin{equation}
\begin{array}{lll}
  I_{1}^{M1} = -i\sqrt{\pi}\cos{(\theta)}, &
  I_{2}^{E1} = -i\sqrt{\displaystyle\frac{2\pi}{3}}, &
  I_{3}^{E1} = -i\sqrt{\displaystyle\frac{\pi}{12}}
                (1 - 3\cos^{2}{(\theta)}),
\end{array}
\label{eq.3.2.14}
\end{equation}                                          %       (3.2.14)
\begin{equation}
  Z_{\rm eff} =
        \displaystyle\frac{2A_{d} - 4Z_{d}}{A_{d} + 4},
\label{eq.3.2.15}
\end{equation}                                          %       (3.2.15)
where $A_{d}$ and $Z_{d}$ are the mass number and charge of the
daughter nucleus, and $\theta$ is the angle between the direction of
the $\alpha$-particle propagation and the direction of the photon
emission.
% ***************************************************************************

% ***************************************************************************
%       The emission spectrum in the alpha-decay of
%       the nucleus 210 Po
% ***************************************************************************
\section{Angular analysis of the bremsstrahlung spectrum in 
$^{210}_{84}$Po $\alpha$-decay}
\label{sec.4}

\begin{figure}[t]
%       Original variant: for sending into Progr. of Theor. Phys.
\centerline{\includegraphics[width=10cm]{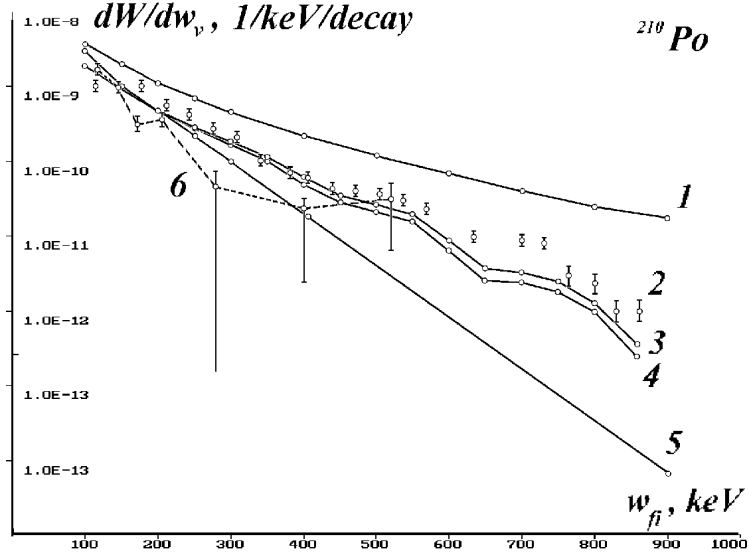}}
\caption{
Bremsstrahlung in the $\alpha$-decay of $^{210}_{84}$Po:
1 --- the result of Tkalya model \cite{Tkalya.1999.PHRVA} based on classical
electrodynamics;
2 --- experimental data;\cite{Eremin.2000.PRLTA}
3 --- the result of our model for $90^{\circ}$ with but E1 and
sub-barrier photon emission;
4 --- the result of our model for $90^{\circ}$ with E1 but without
sub-barrier photon emission;
5 --- the result of Tkalya quantum mechanical model taking into account E1
\cite{Tkalya.1999.PHRVA}
and that of Papenbrock and Bertsch model \cite{Papenbrock.1998.PRLTA};
6 --- experimental data \cite{Kasagi.1997.JPHGB}.}
\label{fig.1}
\end{figure}

We calculated the bremsstrahlung spectra for the $\alpha$-decay of 
$^{210}_{84}$Po at angular values $25^{\circ}$ and $90^{\circ}$
using the model described above.
Our results for $90^{\circ}$ are presented in Fig.~\ref{fig.1}.
In the calculations, we used the parameter values of the Coulomb
barrier from Ref.~\citen{Papenbrock.1998.PRLTA} and the wave function of the
initial $i$-state as the decay state and the final $f$-state as the
scattered state.

In Fig.~\ref{fig.1}, it can be seen that the line 3 obtained using
our model for $90^{\circ}$ lies nearer the experimental
data\cite{Eremin.2000.PRLTA} than both the line 5 representing the
bremsstrahlung spectra obtained using the models of Papenbrock and
Bertsch\cite{Papenbrock.1998.PRLTA} and Tkalya\cite{Tkalya.1999.PHRVA} (taking
into account E1), based on quantum electrodynamics, and line 1
representing the bremsstrahlung spectrum obtained using Tkalya
model\cite{Tkalya.1999.PHRVA} based on classical electrodynamics.
Such an accurate description of the experimental
data\cite{Eremin.2000.PRLTA} of the bremsstrahlung spectrum for the
nucleus $^{210}$Po has been obtained using a multipole approach
without the transformation (\ref{eq.3.2.2}).

The total bremsstrahlung spectrum for $^{210}$Po calculated for
$90^{\circ}$ using our model is represented by line 3 in
Fig.~\ref{fig.1} and is seen to be monotonic. The spectrum calculated
for $25^{\circ}$ lies close to it. With angular analyzing, our model
is found to predict a monotonic dependence of spectra on photon
energy for any angle value (in the region $5^{\circ}$ -- $90^{\circ}$)
and does not yield a slope as large as that of the experimental
data\cite{Kasagi.1997.JPHGB}. Taking into account M1, E2 and M2
multipoles the monotonic behaviour of the bremsstrahlung spectra
for any angular value exists also where there is no visible ``hole''
with such amplitude, as in the experimental data\cite{Kasagi.1997.JPHGB}.
As we have seen, the quantum mechanical models of Papenbrock,
Bertsch\cite{Papenbrock.1998.PRLTA} and Tkalya\cite{Tkalya.1999.PHRVA} also
fail to explain the existence of the ``hole'' in the spectrum.

% A monotonic spectrum is also found in calculations where the
% mixed region is analyzed \cite{Takigawa.1998.PHRVA}.
%
Note that taking account of the mixed region \cite{Takigawa.1998.PHRVA} is
useful for detailed analysis of photon emission from different
regions and the contributions to the total bremsstrahlung spectrum.
However, we find that the total bremsstrahlung spectrum, calculated
on the basis of the quantum mechanical models \cite{Papenbrock.1998.PRLTA,
Tkalya.1999.PHRVA} and our model, is not depended on the choice of
the mixed region and its change.

In addition, we calculated the bremsstrahlung spectrum for
$^{210}$Po without taking into account the sub-barrier emission of
the photons. The resulting spectrum is represented by line 4 in
Fig.~\ref{fig.1} for $90^{\circ}$. Here, it can be seen that line 4
lies somewhat farther from the experimental data\cite{Eremin.2000.PRLTA}
(obtained for $90^{\circ}$ also) than line 3 for the total
bremsstrahlung spectrum. From this we can conclude that taking
account of the photon emission from the barrier region in the
$\alpha$-decay of $^{210}$Po increases the accuracy of experimental
spectra description. But sub-barrier bremsstrahlung gives a small
contribution to the total bremsstrahlung spectrum.
% ***************************************************************************

% ***************************************************************************
\section{Conclusions and perspectives}
\label{sec.7}

We presented a quantum mechanical approach to the calculation of the
brems-\linebreak strahlung spectrum in the $\alpha$-decay of heavy
nuclei, where the angle between the directions of the
$\alpha$-particle propagation and the photon emission is taken into
account. On the basis of the spectrum calculation carried out with
this approach, we conclude that the model gives a monotonic
bremsstrahlung spectrum for the different angles.
(For readers interesting in the comparative analysis of the
theoretical or experimental bremsstrahlung spectra obtained for
different angles, we can also propose a new method presented
in~\citen{Maydanyuk.2004.PAST.brem},
where the dependence of the bremsstrahlung spectra on the values
of such angle is obtained in a simple analytical form and is more
obvious and simple.)

According with analysis by our model, taking into account multipoles
of larger orders does not cause the spectra to become non-monotonic
and does not explain the existence of the ``hole'' in the
experimental data~\citen{Kasagi.1997.JPHGB}.
% As we have seen, the quantum mechanical models of Papenbrock and
% Bertsch\cite{Papenbrock.1998.PRLTA} and Tkalya\cite{Tkalya.1999.PHRVA} also fail
% to explain the existence of the ``hole'' in the spectrum.
The ``hole'' in the total bremsstrahlung spectrum reported in
Ref.~\citen{Kasagi.1997.JPHGB} cannot be explained by the choice of
$R_{2}$, its adjustment, the arrangement of the mixed region (which
was introduced in Ref.~\citen{Takigawa.1998.PHRVA}) nor use of the isotropic
quantum mechanical model, where the potential starting from $R_{1}$
is pure Coulomb. (Here, $R_{1}$ and $R_{2}$ are the internal and
external radii of the Coulomb
barrier\cite{Maydanyuk.2001.UPJ,Tkalya.1999.PHRVA,Takigawa.1998.PHRVA}.)
We agree with Tkalya that it would be useful to obtain more
detailed experimental data for the bremsstrahlung spectrum in the
$\alpha$-decay of $^{210}$Po for different angle values.

From a comparative analysis,  it is found that taking into account
the sub-barrier bremsstrahlung phenomena in $\alpha$-decay of
the nucleus $^{210}$Po increases the accuracy of experimental spectra
description. But sub-barrier bremsstrahlung gives a small
contribution to the total bremsstrahlung spectrum (see
Fig.~\ref{fig.1}). From this result, we can explain the fact that
the models constructed on the basis of classical electrodynamics and
those employing a semiclassical approximation describe the
experimental data for the bremsstrahlung spectra of other nuclei
sufficiently well.

We would also like to note further matters regarding the investigation
of sub-barrier bremsstrahlung phenomena.
It would be useful to obtain more accurate experimental data for the
nucleus $^{210}$Po for photons energies $w_{fi} \geq 400~\mbox {MeV}$
and for the different angle values.
The accuracy of the description of the $\alpha$-decay with photon
emission can be improved by accounting for the individual properties
of the nucleus.
According to our model, small oscillations exist in the
bremsstrahlung spectrum for $^{210}$Po and the period can be
estimated. The existence of oscillation can be explained through use
of the multipole approach without recourse to the transformation
(\ref{eq.3.2.2}) in the calculation of the bremsstrahlung spectrum.
% ***************************************************************************

% ***************************************************************************
\section*{Acknowledgements}
The authors are grateful to Dr.~J.~Jakiel for productive discussions
concerning the study of sub-barrier bremsstrahlung phenomena and
the calculations of the tunneling times, and
to Dr.~A.~P.~Undinko for the valuable suggestions in the calculations
of Coulomb functions with higher accuracy.
% ***************************************************************************

% ***************************************************************************
%                               Appendix A
% ***************************************************************************
\appendix
\section{Calculation of Multipoles}
\label{app.1}

In calculating the bremsstrahlung spectrum, we use a matrix element
of the form
\begin{equation}
  \langle f | \boldsymbol{p} e^{i \boldsymbol{kr}} | i \rangle.
\label{eq.app.1}
\end{equation}                                          %       (app.1)
In the calculation of this quantity, Tkalya\cite{Tkalya.1999.PHRVA} and
Papenbrock and Bertsch\cite{Papenbrock.1998.PRLTA} used the following
transformation:
\begin{equation}
  -\langle f | \boldsymbol{p} | i \rangle =
  \langle f |[H, \boldsymbol{p}]| i \rangle \displaystyle\frac{1}{w_{fi}} =
  i\hbar \biggl\langle f \biggl|
  \displaystyle\frac{\partial}{\partial \boldsymbol{r}} U(\boldsymbol{r}) \biggr|
  i \biggl \rangle \displaystyle\frac{1}{w_{fi}}.
\label{eq.app.2}
\end{equation}                                          %       (app.2)
However, this transformation can be used only in case that 
approximation 
\begin{equation}
  e^{i\boldsymbol{kr}} = 1
\label{eq.app.3}
\end{equation}                                          %       (app.3)
 is made. 
Indeed, taking into account the properties
%
%\[
\begin{equation}
%\begin{array}{l}
  H f(\boldsymbol{r}) \psi(\boldsymbol{r}) =
  \biggl(\displaystyle\frac{\hbar^{2}}{2m}\Delta + U(\boldsymbol{r})\biggr)
  f(\boldsymbol{r}) \psi(\boldsymbol{r}) \neq
%  \neq
  f(\boldsymbol{r}) \biggl(\displaystyle\frac{\hbar^{2}}{2m}\Delta +
  U(\boldsymbol{r})\biggr) \psi(\boldsymbol{r}) =
  f(\boldsymbol{r}) H \psi(\boldsymbol{r}),
%\end{array}
\label{eq.app.4}
\end{equation}                                          %       (app.4)
%\]
%
we have
%\[
\begin{eqnarray}
  \langle f | H\boldsymbol{ p} f(\boldsymbol{r}) | i \rangle =
  \langle f | E_{f}\boldsymbol{p} f(\boldsymbol{r}) | i \rangle =
  E_{f} \langle f | \boldsymbol{p} f(\boldsymbol{r}) | i \rangle,
  \nonumber   \\
  \langle f | \boldsymbol{p} H f(\boldsymbol{r}) | i \rangle \neq
  \langle f | \boldsymbol{p} f(\boldsymbol{r}) H | i \rangle =
  \langle f | \boldsymbol{p} f(\boldsymbol{r}) E_{i}| i \rangle =
  E_{i} \langle f | \boldsymbol{p} f(\boldsymbol{r}) | i \rangle.
\label{eq.app.5}
\end{eqnarray}                                          %       (app.5)
%\]
Only in the case $f(\boldsymbol{r}) = {\rm const}$ does the
inequality (\ref{eq.app.4}) become an equality, in which case the
transformation (\ref{eq.app.2}) can be used to simplify the
calculation of (\ref{eq.app.1}). In a multipole expansion, we have
%
%\[
\begin{equation}
  f(\boldsymbol{r}) \to
  e^{i \boldsymbol{kr}} =
  4\pi \sum\limits_{l=0}^{+\infty} \sum\limits_{m=-l}^{m=l}
  i^{l} j_{l}(kr) Y_{lm}^{*}\biggl(\displaystyle\frac{\boldsymbol{k}}{k}\biggr)
  Y_{lm}\biggl(\displaystyle\frac{\boldsymbol{k}}{k}\biggr),
\label{eq.app.6}
\end{equation}                                          %       (app.6)
%\]
%
where
$j_{l}(kr)$ is the spherically symmetric Bessel function of order $l$.
% ***************************************************************************

% ***************************************************************************
% Some macros are available for the bibliography:
%   o for general use
%      \JL : general journals          \andvol : Vol (Year) Page
%   o for individual journal 
%    \AJ   : Astrophys. J.           \NC         : Nuovo Cim.
%    \ANN  : Ann. of Phys.           \NPA, \NPB  : Nucl. Phys. [A,B]
%    \CMP  : Commun. Math. Phys.     \PLA, \PLB  : Phys. Lett. [A,B]
%    \IJMP : Int. J. Mod. Phys.      \PRA - \PRE : Phys. Rev. [A-E]     
%    \JHEP : J. High Energy Phys.    \PRL        : Phys. Rev. Lett.
%    \JMP  : J. Math. Phys.          \PRP        : Phys. Rep.
%    \JP   : J. of Phys.             \PTP        : Prog. Theor. Phys.     
%    \JPSJ : J. Phys. Soc. Jpn.      \PTPS       : Prog. Theor. Phys. Suppl
% Usage:
%   \PR{D45,1990,345}          ==> Phys.~Rev.\ \textbf{D45} (1990), 345
%   \JL{Nature,418,2002,123}   ==> Nature \textbf{418} (2002), 123
%   \andvol{B123,1995,1020}    ==> \textbf{B123} (1995), 1020
%---------------------------------------------------------------------------

% \bibliographystyle{utphys}
\bibliographystyle{h-physrev4}
% \bibliographystyle{h-physrev}
% \bibliographystyle{JHEP-2}

% \bibliography{2904}

\end{document}